\documentclass[referee]{raa}            % referee version: for submission

\usepackage{graphicx,times}             %for PS/EPS graphics inclusion, new

\usepackage{amssymb,amsmath}

\usepackage{natbib}
\bibpunct{(}{)}{;}{a}{}{,}

\newcommand{\nodenunits}{~$\rm cm^{-3}$}

\newcommand{\adsr}{    {Adv. Space Res.}}
\newcommand{\angeo}{   {Annales Geophysicae}}

\begin{document}
   \title{Interpretation of the coronal magnetic field configuration of the Sun
%\,$^*$
%\footnotetext{$*$ Supported by the National Natural Science Foundation of China.}
}
%   \subtitle{I. Place Your Subtitle Here}

   \volnopage{Vol.0 (200x) No.0, 000--000}      %%preserved for Editor. DOn't remove!
   \setcounter{page}{1}          %%starting page, preserved for Editor. DOn't remove!

   \author{Bo Li \inst{1,2} \and Xing Li \inst{3} \and Hui Yu\inst{1}
   }

   \institute{Shandong Provincial Key Laboratory of Optical Astronomy \& Solar-Terrestrial Environment,
            School of Space Science and Physics,
            Shandong University at Weihai, Weihai 264209, China {\it bbl@sdu.edu.cn} \\
        \and
             State Key Laboratory of Space Weather, Chinese Academy of Sciences, Beijing 100190, China \\
        \and
             Institute of Mathematics \& Physics, Aberystwyth University, Aberystwyth SY23 3BZ, UK
   }

   \date{Received~~2012 04 day; accepted~~year~~month day}

\abstract{
The origin of the heliospheric magnetic flux on the Sun, and hence the origin of the solar wind, is a topic of hot debate.
%This is the case even for solar minimum conditions.
While the prevailing view is that the solar wind originates from outside coronal streamer helmets,
   there also exists the suggestion that the open magnetic field spans a far wider region.
Without the definitive measurement of the coronal magnetic field,
   it is difficult to resolve the conflict between the two scenarios without doubt.
We present two 2-dimensional, Alfv\'enic-turbulence-based models of
   the solar corona and solar wind, one with and the other without a
   closed magnetic field region in the inner corona.
The purpose of the latter model is to test whether it is possible to realize a picture
   suggested by polarimetric measurements of the corona using the Fe
   XIII 10747\AA\ line, where open magnetic field lines seem to
   penetrate the streamer base.
The boundary conditions at the coronal base are able to account for important
   observational constraints,
   especially those on the magnetic flux distribution.
Interestingly, the two models provide similar polarized brightness (pB) distributions
   in the field of view (FOV) of
   SOHO/LASCO C2 and C3 coronagraphs.
In particular, a dome-shaped feature is present in the C2 FOV even for the model
   without any closed magnetic field.
Moreover, both models fit equally well the Ulysses data scaled to 1~AU.
We suggest that: 1) The pB observations cannot
   be safely taken as a proxy for the magnetic field topology, as often
   implicitly assumed.
   2) The Ulysses measurements, especially the one showing a nearly uniform distribution with heliocentric latitude of
   the radial magnetic field, do not rule out the ubiquity of open magnetic fields on the Sun.
}

   \authorrunning{B. Li}                        %author_head in even pages
   \titlerunning{Magnetic Topology of Solar Corona}     % title_head in odd pages

   \maketitle

%% The author head (on even pages) and the title head (on odd pages) will be
%% automatically extracted from \author{} and \title{}. Whenever the title is too long,
%% you will be asked to supply a shorter one by inserting either \authorrunning{} or
%% \titlerunning{} before \maketitle. Anyway, you can specify your own heads.
%%
%%
%% Note: In the following text body of your manuscript, please note several differences from
%%       other major journals:
%% (1) \subsection{Please Capitalize the First Letter of Each Notional Word in Subsection Title}
%% (2) Please Capitalize the First Letter of Each Notional Word in all tables' captions

%
%________________________________________________ sections below
%

\section{Introduction}
\label{sec_intro}
Identifying the source regions of the heliospheric magnetic flux, and hence
    those of the solar wind, is a long standing issue
    in solar physics~\citep[see e.g.,][]{Schwenn_06,YMWang_09}.
The difficulties associated with this identification are due mainly
    to the difficulty of directly measuring the solar coronal magnetic
    field, which is essential in most of the schemes mapping the in situ
    solar winds to their coronal sources~\citep[][to name but a
few]{Levine_etal_77, WangSheeley_90, Neugebauer_etal_98, Neugebauer_etal_02, WangSheeley_06, Abbo_etal_10}.
Although advances on polarimetric measurements with coronal emission lines have been
    recently made~\citep{Habbal_etal_01, Habbal_etal_03, Lin_etal_04, YLiu_09},
    the coronal magnetic field remains largely unknown~\citep[e.g.,][]{Cargill_09}.

Without detailed, quantitative measurements, the coronal magnetic field has been commonly
    constructed via numerical extrapolation.
While all available schemes use the photospheric magnetic field as boundary input,
    they differ substantially in how to treat the effects of electric currents on the global coronal magnetic field
    in a volume bounded by the photosphere and some outer boundary.
The electric currents may be neglected altogether~\citep[e.g., the
     potential field source surface model by][]{Schatten_etal_69, AltschulerNewkirk_69}, they may be assumed to be purely horizontal
     (e.g., the current sheet source surface model
       by~\citeauthor{ZhaoHoeksema_95}~\citeyear{ZhaoHoeksema_95}),
     or flow exclusively along magnetic field lines
     (e.g., the force-free model by~\citeauthor{Tadesse_etal_09}~\citeyear{Tadesse_etal_09}),
     or both the volumetric and sheet currents are self-consistently computed as a product of
      the plasma properties (the
      magnetohydrodynamic (MHD) models by e.g.,~\citeauthor{Lionello_etal_09}~\citeyear{Lionello_etal_09}).
Polarized brightness (pB) images of the solar corona,
      routinely obtained with space-borne and ground-based coronagraphs,
      often guide the extrapolation schemes such that
      the resultant magnetic field configuration matches
      available pB images.
Implied here is that the density structures, as manifested in pB images,
      reflect the magnetic topology in the inner corona.
And usually dome-shaped streamer helmets, the most prominent feature in pB images,
      are seen as comprising closed magnetic fields.
It follows that the bulk of the solar wind originates from
      open field regions outside streamer helmets, even though the
      precise fraction by which coronal holes and the quiet Sun contribute
      to the solar wind is debatable~\citep[][and references therein]{Kopp_94, Hu_etal_03}.

However, this scenario is not universally accepted.
A distinct picture has been advocated in which the solar wind flows
    along the ubiquitous open magnetic field lines that are not limited
    to coronal holes or the quiet Sun but come from throughout the
    Sun~\citep{WooHabbal_99, WooHabbal_03, Woo_etal_04}~\citep[see also][]{WooDruck_08}.
Interestingly, the arguments raised to support this picture initially also came from density
    measurements.
By combining the near-Sun pB values with radio occultation measurements as well as in situ solar wind data,
    \citet{WooHabbal_99} showed that the signatures of
    coronal sources are preserved in the measured solar wind
    away from the Sun, contending that these density imprints are almost
    radially propagating.
A further support for this scenario comes from the fact that the white
    light images of the Sun at total eclipses, when properly processed, exhibit
    a rich set of filamentary structures that extend almost radially from the solar
    surface~\citep[see Fig.1 in ][]{WooHabbal_03}.
Supposing these fine structures trace the magnetic field lines, this would suggest that
    some coronal magnetic field lines penetrate the dome-shaped streamer base.
More importantly, these apparently open magnetic fields were also seen in the polarimetric
    measurements of the inner corona using
    Fe XIII 10747\AA\ line~\citep[see figures in][]{Habbal_etal_01, Habbal_etal_03}.
Carefully addressing observational complications such as collisional depolarization and the Van Fleck effect,
    \citet{Habbal_etal_01} argued that the largely radially aligned
    polarization vectors may indeed reflect a coronal magnetic field that is predominantly radial.

Given the importance of addressing the origin of the open magnetic
    flux of the heliosphere, it is surprising to see that while the traditional scenario
    has been extensively incorporated into numerical studies originated
     by~\citet{PneumanKopp_71}~\citep[also see][and references therein]{Lionello_etal_09},
    the scenario proposed by~\citet{WooHabbal_03}
    has not been modeled quantitatively.
Here we wish to implement this scenario in a numerical model, thereby
    testing it against two fundamental constraints that the traditional
    scenario can readily satisfy:
    one is the appearance of a
    dome-shaped bright feature in pB images, and the other is the fact
    established by Ulysses measurements that the radial magnetic field
    strength $B_r$ is nearly uniform with latitude
    beyond 1~AU~\citep{SmithBalogh_95, Smith_etal_01}.
Note that the latter fact was used to refute the suggestion of a largely radially
    expanding solar wind~\citep{Smith_etal_01},
    as the magnetic flux near the Sun is obviously nonuniform~\citep[e.g.,][]{Svalgaard_78, Vasquez_etal_03}.
Before proceeding, we note that from SOHO/EIT images, it is obvious that
    there are a myriad of low-lying loop-like structures in the corona.
As suggested by~\citet{Habbal_etal_01}, a large-scale closed magnetic field
    (the ``nonradial'' component in their paper) may also
    help shape the large-scale corona.
To simplify our treatment, we will simply try to answer one question:
    can a dome-shaped bright feature show up in a magnetic configuration
    where there is no closed field at all?
In essence, this is equivalent to saying that we are interested in the region
    somewhere above the layer below which loops abound in SOHO/EIT images, and
     beyond which the contribution from closed magnetic fields is assumed to be negligible.

In what follows, we will present two numerical models that differ in the magnetic field
    configuration in the inner corona, one with and
    the other without a closed field region.
Both models are able to incorporate the essential observational
     constraints near the coronal base, especially the latitudinal
     dependence of the radial magnetic field.
We then construct pB maps to see whether they display features
     similar to what is seen in white light observations.
Moreover, model results are also compared with several crucial parameters
     observed in situ.
The models are described in section~\ref{sec_nummodel}, results from the numerical computations
     are given in section~\ref{sec_numresults}, and section~\ref{sec_conc} concludes this paper.

\section{Description of the Numerical Model}
\label{sec_nummodel}

heiti
Assuming azimuthal symmetry, our models solve in the meridional plane ($r$, $\theta$) the standard
    two-fluid MHD equations identical to those in~\citet{Hu_etal_03}.
Here $r$ is the heliocentric distance, and $\theta$ is the colatitude.
The protons are heated by a flux of Alfv\'{e}n waves dissipated at the Kolmogorov
   rate $Q_{\mathrm{kol}}=\rho\left<\delta v^2\right>^{3/2}/L_c$,
   where $\left<\delta v^2\right>^{1/2}$ is the rms amplitude of
   velocity fluctuations of the wave field,
   $\rho= n m_p$ is the mass density in which $n$ is the number density and $m_p$ the proton mass.
The dissipation length $L_c$ is given by $L_c=g\left(\psi\right) L_{0}\left(B_{b,p}/B\right)^{1/2}$ where $B$
   denotes the magnetic field strength, the subscript $b$ represents the value at the coronal base, $B_{b,p}$
    and $L_0$ are respectively $B$ and $L_c$ evaluated at the pole ($r=1$~$R_\odot$ and
   $\theta=0^\circ$).
In addition, $g(\psi)$ describes the dependence of $L_c$ on $\psi$, the magnetic flux function which labels flow tubes.
For the electrons, the classical field-aligned heat conduction is considered.
While not directly heated, electrons receive part of the dissipated wave energy via Coulomb collisions
   with protons, whose heat conduction is neglected.
With properly specified boundary conditions and given $g(\psi)$ as well as $L_0$,
   our numerical computation starts with an arbitrary initial state and
   runs until a steady state is found.
This steady state does not rely on the initial state but is solely determined by $g(\psi)$, $L_0$
   and the boundary conditions at the coronal base.
As a product, the computation yields a global distribution in the $r-\theta$ plane of
   the proton number density $n$, the radial and latitudinal components
   of the proton velocity $v_r$ and $v_{\theta}$, the magnetic flux
   function $\psi$, the electron and proton temperatures $T_e$ and
   $T_p$, as well as the wave pressure $p_w=\rho \left<\delta v^2\right>/2$.
The magnetic field $\vec{B}$ is then obtained via $\vec{B} = \nabla\psi \times \vec{e}_\phi/r\sin\theta$, where
   $\vec{e}_\phi$ represents the unit vector along the azimuthal direction.

The numerical scheme has been described in detail by~\citet{Hu_etal_03}.
The computational domain extends from the coronal base (1~$R_\odot$) to 1~AU, and
     from the pole (colatitude $\theta=0^\circ$) to equator ($\theta=90^\circ$).
Both the pole and equator are taken as symmetrical boundaries.
At the top boundary (1~AU), all dependent variables are linearly
   extrapolated for simplicity.
On the other hand, at the coronal base, the proton density $n$, magnetic flux $\psi$, temperatures ($T_e$
   and $T_p$) and wave pressure $p_w$ are all fixed.
The velocity components $v_r$ and $v_\theta$ are derived from the requirements
   that $\vec{v}$ is aligned with $\vec{B}$, and mass flux
   is conserved along magnetic field lines.
What distinguishes the two models lies in how the values for $(n, T_e, T_p, \psi, p_w)$
   are imposed at the coronal base, and also in how $g(\psi)$ is prescribed.
To proceed, in what follows let the two models be labeled FO (fully open) and PO (partially open),
   respectively.

\begin{figure}
\centering
\includegraphics[angle=90,width=.8\textwidth]{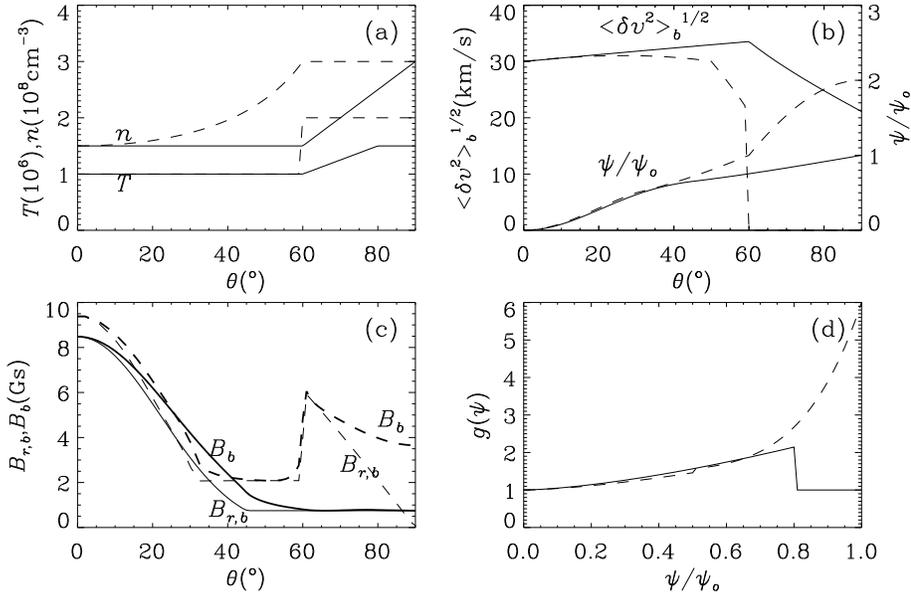}
\caption{Boundary conditions imposed at the coronal base.
   (a): Electron density $n$ and temperature $T$.
        Electrons and protons are assumed to have identical temperatures at the base.
   (b): Wave amplitude $\left<\delta v^2\right>^{1/2}$ and magnetic flux $\psi$ in units of
            the open flux $\psi_o$.
   (c): Radial magnetic field $B_r$ and the total field strength $B$.
   (d): Behavior of the function $g\left(\psi \right)$ that dictates the wave dissipation length
        (please see text).
   Model FO is given by the continuous curves while model PO by the dashed ones.
}
\label{fig-bc}
\end{figure}

Figure~\ref{fig-bc} shows the distribution at the coronal base with
     colatitude $\theta$ of (a): the density $n$ and temperature $T$ (the
     protons and electron ones are assumed to be equal at the base),
     (b): the wave amplitude $\left<\delta v^2\right>^{1/2}$ and magnetic flux given by
          $\psi/\psi_o$, $\psi_o$ being the total open flux,
     (c): the radial magnetic field $B_{r}$ (thin curves) and  magnetic field strength $B$ (thick lines).
Figure~\ref{fig-bc}d describes the behavior of $g(\psi)$.
The solid lines are for model FO whereas the dashed lines describe model PO.
For model PO, the choice of the latitudinal profile for $(n, T_e, T_p, \psi, p_w)$
     and that of $g(\psi)$ have been described in
    considerable detail in~\citet{Hu_etal_03}, and therefore we put more
    emphasis on the justification of model FO.
Let us start with the distribution of the magnetic flux function $\psi$.
Both computations adopt the same value for the open
    flux $\psi_o=7.37\times 10^{21}$~Mx, which corresponds to an average radial magnetic field
    of $3.3\gamma$ at 1~AU, compatible with Ulysses
    measurements~\citep{SmithBalogh_95, Smith_etal_01}.
Figure~\ref{fig-bc}b indicates that in model PO the total flux ($\psi$ at equator)
    is assumed to be twice the open flux, and only
    the portion $\theta \le 60^\circ$ of the solar surface contributes
    to the open flux.
However, in model FO the interplanetary magnetic flux~$\psi_o$ is assumed to emerge from all over the Sun, evidenced
    by the fact that the magnetic flux $\psi$ at equator equals the
    total open flux $\psi_o$.
Actually, the distribution of $\psi$ with $\theta$ derives from the $\theta-$profile of $B_r$ at the coronal
    base, presented by the thin curves in Fig.\ref{fig-bc}c.
The specific form of $\psi$ adopted in FO reflects the observed fact that
    $B_{r,b} \propto \cos^{7}\theta$ for $\theta \lesssim 45^\circ$
    and is roughly constant elsewhere, established by the measurements
    from Kitt Peak synoptic magnetograms at solar
    minima~\citep[see][Fig.3]{Vasquez_etal_03}.

Figure~\ref{fig-bc}a indicates that the two models do not differ
    substantially in the profiles of $n$ and $T$.
While model PO uses a step function in $T$ with $T$ jumping from $1$~MK to $2$~MK at
    $\theta=60^\circ$, model FO adopts a somehow smoother distribution
    of $T$ with a ramp connecting two values of $1$ and $1.5$~MK, the
    ramp spanning the range $60^\circ \le \theta\le 80^\circ$.
It suffices to note that, the quoted value of $1$~MK is compatible with electron temperatures
    in coronal holes, and values of $1.5$ and $2$~MK are in line with electron temperatures in the quiet Sun and
    inside the streamer base~\citep{Habbal_etal_93, JLi_etal_98}.
For the $n$ profile, model PO uses a distribution which is uniform in
    the range $60^\circ \le \theta\le 90^\circ$, but decreases with
    decreasing $\theta$ for the rest of the solar surface.
The quoted values of $1.5\times 10^8$~\nodenunits\ at the pole and $3\times
    10^8$~\nodenunits\ at equator both agree with spectroscopic
    measurements~\citep{Habbal_etal_93, JLi_etal_98}.
On the other hand, the wave amplitudes given in Fig.\ref{fig-bc}b for both models are
    below the upper limit derived spectroscopically~\citep{Esser_etal_99}.
We note that in model PO, the waves are assumed to be absent in closed field regions.
Finally, as there is no direct constraint on
    the correlation length of the turbulent Alfv\'en waves, the $g(\psi)$
    curves given in Fig.\ref{fig-bc}d are found by a trial-and-error
    procedure such that the model results to be presented best match observations.

\section{Numerical Results}
\label{sec_numresults}

\begin{figure}
\centering
\includegraphics[angle=90,width=.98\textwidth]{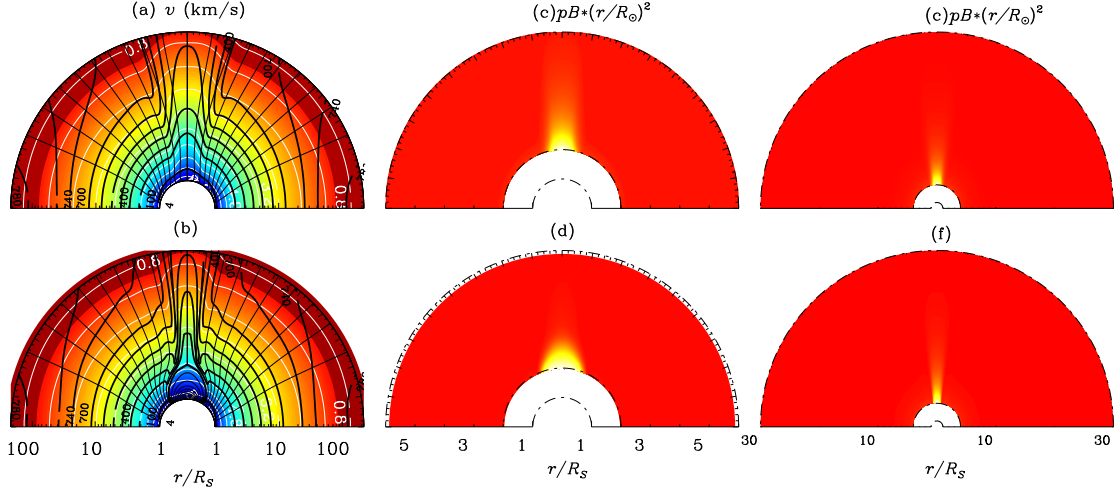}
\caption{
Global solar wind models where the magnetic field is fully open
     (top panels, model FO) and
     partially open (bottom panels, model PO).
(a) and (b): Outflow speed $v$ (thick black contours) superimposed on
     the density map (logarithm of density $\log n$).
     The white curves represent the $\log n$ contours.
     The Background thin black curves are magnetic field lines equally spaced by $0.1 \psi_o$,
     $\psi_o$ being the open flux.
The remaining panels give the polarized brightness (pB) maps computed from the modeled densities $n$.
pB is measured in units of $B_\odot$, the solar mean disk brightness.
Note that ${pB} (r/R_\odot)^2$ is plotted instead of pB.
Panels (c) and (d) are for the heliocentric range of $1-6 R_\odot$,
    while (e) and (f) are for $4-30~R_\odot$.
Note that the two regions correspond to the Field of View for LASCO C2 and C3, respectively.
}
\label{fig-pb}
\end{figure}

The numerical results for models FO and PO are presented in the top and bottom panels of Figure~\ref{fig-pb}, respectively.
Figures~\ref{fig-pb}a and \ref{fig-pb}b give the contours of the outflow speed $v$ (thick black curves) superimposed
    on the distribution of the density $n$.
Moreover, the white contours are for $\log n$, and the background thin curves
    represent the magnetic field lines equally spaced by $0.1\psi_o$.
The polarized brightness maps (in units of the mean disk brightness $B_\odot$ of the Sun) are
    displayed for the heliocentric range of $2-6 R_\odot$ (Figures~\ref{fig-pb}c and~\ref{fig-pb}d)
    and $4-30 R_\odot$ (Figures~\ref{fig-pb}e and~\ref{fig-pb}f),
    corresponding to the Field of View (FOV) of LASCO C2 and C3 coronagraphs on board SOHO.
These pB data are computed from the modeled densities using the standard formulae
    given in e.g., \citet{vandehulst_50, Vasquez_etal_03}.
It is obvious from Figs.\ref{fig-pb}a and~\ref{fig-pb}b that both models produce
    an outflow field characterized by a narrow wedge of slow and dense
    wind ($v \le 400$~km/s) embedded in the fast wind.
In model PO, a region associated with closed magnetic field can be seen whose tip
    is located at around $3.5~R_\odot$.
However, model FO does not possess such a region, the solar wind flows from throughout the Sun.
Despite this, both models produce similar pB maps in the two different heliocentric ranges.
In particular, Figs.\ref{fig-pb}c and~\ref{fig-pb}d both display a dome-like bright feature.
While in model PO this feature is associated with the trapping of plasmas in
    closed field regions, in model FO it is due to the fact that the
    density in the near-equator region decreases more slowly with radial
    distance than $n$ does in the polar region.
Moreover, the high-latitude magnetic filed lines can be seen to bend towards the
    equator to relax the latitudinal gradient in the magnetic pressure.
From this we conclude that, even though the magnetic field lines in model FO are not
    purely radial, the surprising linear polarization measurements
    with Fe XIII 10747\AA\ line placed in the context of pB
    measurements~(e.g., the west limb in Fig.1 of \citeauthor{Habbal_etal_01}~\citeyear{Habbal_etal_01})
    can be understood: the large-scale bright features may not reflect
    closed field regions but are associated with open field lines that
    penetrate them.

\begin{figure}
\centering
\includegraphics[angle=90,width=.9\textwidth]{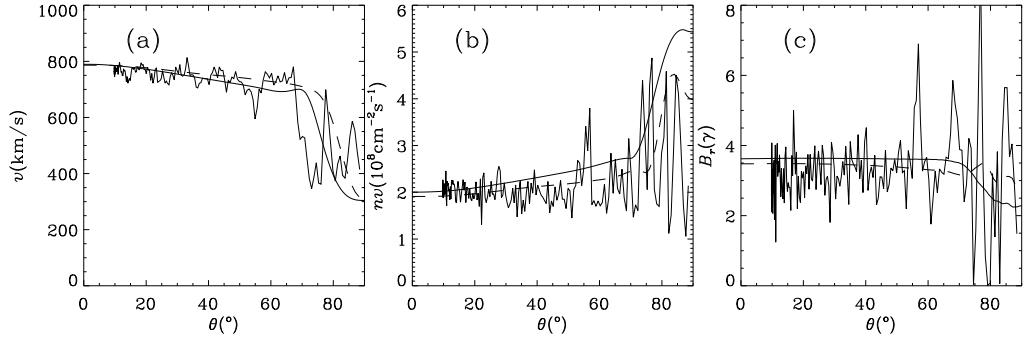}
\caption{
Comparison with Ulysses data of derived latitudinal distributions of various parameters.
(a): Outflow speed $v$. (b): Flux density $nv$.
(c): Radial component of the magnetic field $B_r$.
Model FO is given by the continuous lines, whereas model PO by the dashed lines.
Moreover, in situ measurements by Ulysses are given by the light continuous lines.
}
\label{fig-au}
\end{figure}

Figure~\ref{fig-au} compares the latitudinal distribution at 1~AU of
     (a) the outflow speed $v$,
     (b) flux density $n v$ and
     (c) the radial magnetic field $B_r$, derived from the two models.
Model PO (FO) is described by the dashed (solid) curves.
Also plotted are the daily averages of Ulysses data during the
      first half of the fast latitudinal scan from Sep 12 1994 to
      Mar 4 1995.
The two models, although having distinct magnetic configurations in the
      inner corona, provide equally good fits to the Ulysses data.
One may say that the proton flux density $n v$ in model FO seems to be
      slightly larger than the measured values near the equator, it is
      nonetheless within the ranges typically measured in situ~\citep[Fig.4 in][]{McComas_etal_00}.
Concerning the distribution of $B_r$, model FO actually agrees with the Ulysses measurements better
      than model PO does.
Furthermore, it is interesting to see that $B_r$ hardly displays any latitudinal dependence in the fast wind region.
But this is certainly not the case at the coronal base, shown in Fig.\ref{fig-bc}c.
The relaxation of the latitudinal gradient in $B_r$ can be understood from the force
      balance condition in the direction perpendicular to the ambient magnetic field.
When the magnetic field becomes largely radial, this condition reduces to that $\left(\beta+1\right)B^2/8\pi$ is latitude independent,
      where $\beta=8\pi\left(p_e+p_p+p_w\right)/B^2$, $p_e$ and $p_p$ are
      respectively the electron and proton pressures,
      and $p_w$ the wave pressure.
It turns out that this happens at some distance where $\beta \ll 1$ holds in both models.
Consequently, the magnetic field should exhibit little latitudinal variation from there on.
This relaxation process necessarily leads to a significant non-radial component of the solar coronal magnetic field,
      in view of the base distribution of $B_r$ which varies significantly from the pole to
      equator.
However, a uniform latitudinal distribution of $B_r$ at 1~AU does not necessarily invalidate a scenario where open magnetic
      fluxes originate from throughout the Sun.
In this regard, we agree with \citet{Smith_etal_01} who concluded that the coronal magnetic
      field cannot be exactly radial.

\section{Concluding Remarks}
\label{sec_conc}
Identifying the origin of the Sun's open magnetic
   flux is of crucial importance in establishing the connection of the
   in situ solar winds to their sources.
In the absence of a definitive measurement of the solar coronal magnetic field, this identification
   problem is subject to considerable debate~\citep[e.g.,][]{Schwenn_06, YMWang_09}.
This is true even at solar minima when the configuration of the solar corona is
   relatively simple, the most prominent feature being the bright
   streamer helmets in white light images.
While the prevailing view is that the majority of the solar wind originates from outside streamer
   helmets~\citep{PneumanKopp_71}, there also exists the suggestion that the open magnetic field is
   ubiquitous on the Sun and not confined to coronal holes or the quiet Sun~\citep{WooHabbal_99, WooHabbal_03}.
Implementing the former scenario in a numerical model has been a common practice~\citep[e.g.][]{Lionello_etal_09},
   however, so far the latter has not been modeled quantitatively and hence
   tested quantitatively.
Here we offer such an implementation.

We have constructed two 2-dimensional, Alfv\'enic-turbulence-based models
   of the solar corona and solar wind, one with and the other without a
   closed magnetic field region in the inner corona.
The purpose of the latter model is to minimize the contribution
    of the closed magnetic
   field, trying to mimic a corona permeated with open magnetic fields
   which may infiltrate the dome-shaped streamer base.
In specifying the boundary conditions at the coronal base, we have taken into
   account some important observational constraints, especially those
   on the magnetic flux distributions.
Interestingly, the two models provide similar polarized brightness (pB) maps
   in the field of view (FOV) of the SOHO/LASCO C2 and C3 coronagraphs.
In particular, a dome-shaped feature is present in the C2 FOV even for
   the model without any closed magnetic field.
Moreover, both models fit equally well the Ulysses data scaled to 1~AU.
Hence we suggest that: 1) The pB observations cannot
   be safely taken as a proxy for the magnetic field topology, as often
   implicitly assumed. 2) The Ulysses measurements, especially the one
   indicating that the radial magnetic field strength is nearly
   uniformly distributed with heliocentric latitude, do not rule out
   the ubiquity of open magnetic fields on the Sun.

We do not intend to resolve the conflict of the two distinct scenarios currently
   available for the origin of the heliospheric magnetic flux.
Rather, the presented numerical results suggest the likelihood that the magnetic field structure of bright
   features (e.g., helmet streamers) in the corona may be
   more diverse than traditionally viewed: the magnetic flux therein can be
   either closed or open.
To differentiate the scenarios, it is likely that more stringent constraints
   come from the SOHO/UVCS measurements.
For instance, measurements based on the Doppler dimming technique have
   yielded that along the direction transverse to the streamer helmet, there exists
   a transition in the inferred plasma speed from unmeasurable to significant values,
   and this transition seems to trace the streamer legs~\citep{Strachan_etal_02, Frazin_etal_03},
   identified by the enhancement of the intensity ratio of O VI $\lambda$1032\AA\ to
   H I Ly$\alpha$~\citep{Kohl_etal_97}.
Therefore, it remains to be seen whether a model permeated with open magnetic fluxes can
   account for this feature.
To do this, an obvious need is to incorporate O$^{5+}$ ions in a three-fluid model and test
   both scenarios.
At the moment, in such models only the traditional partially open scenario
   has been adopted~\citep[e.g.,][]{Li_etal_06, Ofman_etal_11}.
Instead of implementing a further construction,
   let us end here by noting that one may also ask whether these observational
   features~\citep{Kohl_etal_97, Strachan_etal_02, Frazin_etal_03} are universal for all streamers.

\begin{acknowledgements}
The Ulysses data are obtained from CDAWeb database.
The two Ulysses teams, SWOOPS (PI: D.J.McComas) and
     VHM (PI: A.Balogh) are gratefully acknowledged.
This research is supported by the National Natural Science Foundation of China (40904047 and 41174154), the Ministry of Education of China
     (20110131110058 and NCET-11-0305), and also by the Specialized Research Fund for State Key Laboratories.
\end{acknowledgements}

\end{document}